\documentstyle[array,aps,twocolumn]{revtex}
\begin{document}
\draft

\wideabs{
  
\title{Self-consistent non-Markovian theory of a quantum state
  evolution for quantum information processing}
  
\author{Doyeol Ahn\thanks{Also with Department of Electrical Engineering,
University of Seoul, Seoul 130-743, Korea}\thanks{Electronic address;
dahn@uoscc.uos.ac.kr}, Jinhyoung Lee, and S. W. Hwang\thanks{Permanent address: Department of Electronics Engineering,
 Korea University, 5-1 Anam, Sungbook-ku, Seoul 136-701, Korea}}
  
\address{Institute of Quantum Information Processing and Systems,
  University of Seoul, 90 Jeonnong, Tongdaemonn-ku, Seoul, Korea}
  
\date{\today}
  
\maketitle

\begin{abstract}
  It is shown that the operator sum representation for non-Markovian
  dynamics and the  Lindblad master equation in Markovian
  limit can be derived from a formal solution to quantum Liouville
  equation for a qubit system in the presence of decoherence processes
  self-consistently. Our formulation is the first principle theory
  based on projection-operator formalism to obtain an exact reduced
  density operator in time-convolutionless form starting from the
  quantum Liouville equation for a noisy quantum computer. The
  advantage of our approach is that it is general enough to describe a
  realistic quantum computer in the presence of decoherence provided
  details of the Hamiltonians are known.
\end{abstract}


}

Dynamics of a quantum system coupled to an environment has been
studied extensively for potential applications to quantum computing
and quantum information processing recently
\cite{Jozsa94,Schumacher96,Lidar98,Peres00}. The key element of the
studies is the reduced-density-operator which is a solution to quantum
Liouville equation (QLE). The QLE would involve Hamiltonians for
systems representing qubits, reservoir, and mutual interaction between
the system and reservoir that causes decoherence \cite{Kim95}.  The
presence of decoherence would be the most important obstacle to the
ideal operations of quantum gates or quantum channels \cite{Lee00}. To
overcome this difficulty, the quantum error correcting codes
\cite{Bennett98} and the decoherence free subspaces \cite{Zanardi97}
of multiple qubit systems have been suggested.  For both quantum error
correcting codes and decoherence free subspaces the knowledge of the
reduced density operator of the qubit system is essential.

Up to now the information about the reduced density operator is
obtained from Lindblad master equation \cite{Lidar98,Lindblad76} in
Markovian approximation or an operator sum representation (OSR)
\cite{Kraus71} in the non-Markovian case which is also known as Kraus
representation. Even though the OSR provides better information about
the qubit system than the Markovian formalism, somewhat surprisingly,
as pointed out by Bacon {\em et al.} \cite{Bacon01}, the former is
obtained in the language of gates, {\em i.e.}, the unitary
transformation, rather than from the solution to the QLE itself in the
Hamiltonian formulation. In other words, most of the proposals for
quantum computers or quantum gates have assumed particular forms of
the unitary transformations beforehand. In our opinion, it would be
desirable if there is a way to obtain the direct solution for the
reduced density operator from the QLE to model physical
implementations of the quantum computers.

The QLE is an integro-differential equation and, in general, it is
nontrivial to obtain the solution of the form
\begin{equation}
  \label{eq:so}
  {\hat \rho} \stackrel{\cal E}{\longrightarrow} {\hat \rho}' = {\cal
  E}[{\hat \rho}]
\end{equation}
where ${\hat \rho}$ is the reduced density operator and $\cal E$ is
the superoperator of linear mapping. The superoperator $\cal E$ is
not necessarily a unitary transformation if one considers an open
system interacting with a reservoir in the presence of decoherence
processes. Sometime ago we studied the time-convolutionless reduced
density operator formulation to model quantum devices
\cite{Ahn94,Ahn97} and noisy quantum channels \cite{Ahn00}. In this
theory the memory kernels of the Volterra-type integral equation are
solved self-consistently using the superoperator formalism and it was
shown that both non-Markovian decoherence process and renormalization
of the memory effects can be incorporated.

In this paper we formulate a general non-Markovian theory based on a
QLE and show that the OSR for the non-Markovian case and the
 Lindblad master equation approach within the Markov
approximation can be derived self-consistently. 

The Hamiltonian of the total system is assumed to be
\begin{eqnarray}
{\hat H}_t(t)={\hat H}_s(t)+{\hat H}_b + {\hat H}_{int},
\end{eqnarray}
where ${\hat H}_s(t)$ is the Hamiltonian of the system, ${\hat H}_b$
the reservoir, and ${\hat H}_{int}$ the Hamiltonian for the
interaction of the system with the reservoir. Note that the system
Hamiltonian ${\hat H}_s(t)$ may contain time-dependent external field
terms to control the qubit system. 
The equation of motion for density operator ${\hat \rho}_t$ of the
total system is given by a QLE as
\begin{eqnarray}
  \label{eq:qle}
  \frac{d}{dt}{\hat \rho}_t(t) = -i[{\hat H}_t(t),{\hat \rho}_t(t)] =
  -i {\cal L}_t(t) {\hat \rho}_t(t), 
\end{eqnarray}
where ${\cal L}_t(t)={\cal L}_s(t) +{\cal L}_b +{\cal L}_{int}$ is the
Liouville operator. The Liouville operators are in one-to-one
correspondence with the Hamiltonians.  Here we use a unit of $\hbar =
1$. The reservoir is assumed to be in the thermal state.
However, the assumption may be extended to any time-independent
reservoir states which commutes with the reservoir Hamiltonian, {\em
  i.e.}, ${\cal L}_b \hat{\rho}_b=0$. In order to derive and to solve
an equation for the system alone, we employ the projection-operators
\cite{Zwanzig60,Saeki82} that decompose the total
system by eliminating the degrees of freedom for the reservoir.
Time-independent projection-operators ${\cal P}$ and ${\cal Q}$ are
defined as
\begin{eqnarray}
{\cal P} {\hat X}= {\hat \rho}_b {\rm Tr}_b({\hat X}),~~
{\cal Q}=1-{\cal P},
\end{eqnarray}
for any dynamical variable ${\hat X}$.  Here ${\rm Tr}_b$ indicates a
partial trace over the quantum reservoir. The information of the
system is contained in the reduced density operator ${\hat \rho}(t)$
given by
\begin{eqnarray}
{\hat \rho}(t) &=& {\rm Tr}_b {\hat \rho}_t(t)\nonumber \\
&=& {\rm Tr}_b {\cal P} {\hat \rho}_t(t).
\end{eqnarray}
After some mathematical manipulations, the time convolutionless
equation of motion for ${\cal P}{\hat
  \rho}_t(t)=\hat{\rho}_b\hat{\rho}(t)$ is given by \cite{Ahn94,Ahn97,Ahn00}
\begin{eqnarray}
\label{eq:peq1}
\frac{d}{dt}{\cal P}{\hat \rho}_t(t)
= -i {\cal P} {\cal L}_t(t) {\cal P} {\hat \rho}_t(t) 
+i {\cal P} {\cal L}_t(t) \left({\cal N}(t)-1\right) {\cal P} {\hat
  \rho}_t(t),   
\end{eqnarray}
where
\begin{eqnarray}
{\cal N}^{-1}(t)=1+i\int_0^t d\tau ~{\cal H}(t,\tau){\cal Q}
{\cal L}_t(\tau){\cal P}~{\cal G}(t,\tau).
\end{eqnarray}
The projected propagator ${\cal H}(t,\tau)$ and the anti-time
evolution operator ${\cal G}(t,\tau)$ of the total system are defined
as
\begin{eqnarray}
{\cal H}(t,\tau)=T\exp\left\{
-i\int_\tau^t ds ~{\cal Q}{\cal L}_t(s){\cal Q}\right\}
\end{eqnarray}
and
\begin{eqnarray*}
{\cal G}(t,\tau)=T^c
\exp\left\{ i \int_\tau^t ds ~{\cal L}_t(s)\right\},
\end{eqnarray*}
where $T$ and $T^c$ denote the time ordering and the anti-time
ordering operators respectively.  The formal solution to
Eq.~(\ref{eq:peq1}) is given by \cite{Ahn00}
\begin{eqnarray}
\label{eq:psol}
{\cal P}{\hat \rho}_t(t)
&=&{\cal U}(t,0){\cal P}{\hat \rho}_t(0) \nonumber \\
& & -i\int_0^t ds ~{\cal U}(t,s){\cal P}{\cal L}_t(s)
\{{\cal N}(s)-1\}{\cal P}{\hat \rho}_t(s),
\end{eqnarray}
where the projected propagator ${\cal U}(t,\tau)$ of the system is
defined by
\begin{eqnarray}
{\cal U}(t,\tau)
=T\exp\left\{
-i\int_\tau^t ds ~{\cal P} {\cal L}_t(s){\cal P}\right\}.
\end{eqnarray}
Eq.~(\ref{eq:psol}) can be put into time-convolutionless form by
substituting
\begin{eqnarray}
{\hat \rho}_t(s)={\cal G}(t,s){\hat \rho}_t(t)
\end{eqnarray}
and after some mathematical manipulations, we obtain the reduced
density operator ${\hat \rho}(t)$, which is an exact solution to the
QLE, given in the form of Eq.~(\ref{eq:so}),
\begin{eqnarray}
\label{eq:rho1}
{\hat \rho}(t)&=&{\cal E}(t){\hat \rho}(0) \nonumber\\
&=& {\cal W}^{-1}(t){\cal U}_s(t,0){\hat \rho}(0),
\end{eqnarray}
with
\begin{eqnarray}
\label{eq:weq}
{\cal W}(t)
&=&1+i\int_0^t ds ~{\cal U}_s(t,s) {\rm Tr}_b\left\{
{\cal L}_{int} {\cal Z}(s)\left(1-{\cal Z}(s)\right)^{-1}{\hat
  \rho}_b\right\} \nonumber \\ 
&&\times{\rm Tr}_b\left\{
{\cal U}_0(s,0){\cal R}(t,s){\cal U}^{-1}_0(t,0)
\left(1-{\cal Z}(t)\right)^{-1}{\hat \rho}_b\right\}.
\end{eqnarray}
Here, we define
\begin{eqnarray}
{\cal Z}(t)&=&1-{\cal N}^{-1}(t), \\
{\cal U}_s(t,\tau)&=&T\exp\left\{
-i\int_\tau^t ds ~{\cal L}_s(s)\right\},\\
{\cal U}_0(t,\tau)&=&\exp\left\{-i(t-\tau) {\cal L}_b\right\}{\cal
  U}_s(t,\tau), \\ 
\end{eqnarray}
and
\begin{eqnarray}
{\cal R}(t,\tau)=
T^c \exp\left\{
i\int^t_\tau ds ~{\cal U}_0^{-1}(s,0){\cal L}_{int} {\cal U}_0(s,0)\right\},
\end{eqnarray}
where ${\cal U}_0(t,\tau)$ is the non-interacting time-evolution
operator of the system and the reservoir and ${\cal R}(t,\tau)$ is the
anti-time evolution operator of the total system in the interaction
picture \cite{Ahn97,Ahn00}.

It is straightforward to obtain the time-convolutionless equation of
motion for a reduced density operator ${\hat \rho}(t)$. From
Eq.~(\ref{eq:peq1}), we get
\begin{equation}
  \label{eq:seq}
\frac{d}{dt}{\hat \rho}(t) = -i {\cal L}_s(t) {\hat \rho}(t) + {\cal
  C}(t) {\hat \rho}(t), 
\end{equation}
with
\begin{equation}
  \label{eq:col}
  {\cal C}(t) = -i {\rm Tr}_b \left\{ {\cal L}_{int} {\cal Z}(t)
  (1-{\cal Z}(t))^{-1} {\hat \rho}_b \right\}
\end{equation}
where ${\cal C}(t)$ is a generalized collision operator and we use an
anzatz ${\cal P}{\cal L}_{int}{\cal P}=0$ which is equivalent to
neglect renormalization of the unperturbed energy of the system
\cite{Saeki82}.

In the following, we first show that the time-convolutionless equation
of motion (\ref{eq:seq}) becomes the  Lindblad master
equation in the Markov approximation. The lowest-order Born
approximation, which is valid up to the order $({\hat H}_{int})^2$, is
used subsequently. The effect of ${\cal C}(t)$ on ${\hat \rho}(t)$ up
to the second-order expansion becomes
\begin{equation}
  \label{eq:socol}
  {\cal C}^{(2)}(t) {\hat \rho}(t) = - \int_0^t d\tau ~{\rm Tr}_b
  \left[{\hat H}_{int}, \left[{\hat H}_{int}(\tau - t),{\hat
  \rho}_b{\hat \rho}(t)\right]\right], 
\end{equation}
where ${\hat H}_{int}(t)$ is the Heisenberg transformation of
${\hat H}_{int}$ defined by ${\cal U}_0(t) {\hat H}_{int}$.
For the specific form of the interaction Hamiltonian, we assume a
Caldeira-Leggett-type model~\cite{Caldeira85,Loss98} given by
\begin{equation}
  \label{eq:cli}
  {\hat H}_{int} = \sum_{\alpha} {\hat v}_\alpha \otimes {\hat
  b}_\alpha 
\end{equation}
where ${\hat v}_\alpha$ is the Hermitian operator acting on the system
and ${\hat b}_\alpha = \sum_k (g_{\alpha k} {\hat a}^\dagger_k +
g_{\alpha k}^* {\hat a}_k)$ is a fluctuating bosonic quantum field
whose unperturbed motion is governed by the harmonic oscillator
Hamiltonian for the reservoir,
\begin{eqnarray}
\label{eq:bath}
{\hat H}_b(t) = \sum_{k} \omega_k {\hat a}^{\dagger}_k {\hat a}_k.
\end{eqnarray}
The set of operators $\{{\hat v}_\alpha\}$ describes the various
decoherence processes and sometimes they are denoted as the error
generators. From Eqs.~(\ref{eq:socol}) and (\ref{eq:cli}), we obtain
\begin{eqnarray}
  \label{eq:socol2}
  {\cal C}^{(2)}(t) {\hat \rho}(t) &=& \sum_{\alpha\beta} \int_0^t d\tau
  ~\chi_{\alpha\beta}(\tau-t) [{\hat v}_\beta(\tau-t){\hat \rho}(t),
  {\hat v}_\alpha] \nonumber \\ 
  & &+ \sum_{\alpha\beta} \int_0^t d\tau
  ~\chi_{\alpha\beta}(t-\tau)[{\hat v}_\alpha,
  {\hat \rho}(t){\hat v}_\beta(\tau-t)] \nonumber \\
\end{eqnarray}
where
\begin{eqnarray}
  \label{eq:chif}
  \chi_{\alpha\beta}(t) = {\rm Tr}_b{\hat b}_{\alpha}(t){\hat
  b}_{\beta}{\hat \rho}_b = {\rm Tr}_b{\hat b}_{\alpha} {\hat
  b}_{\beta}(-t) {\hat \rho}_b. 
\end{eqnarray}
The characteristic function $\chi_{\alpha\beta}(t)$ for the heat bath
satisfies 
$\chi_{\alpha\beta}(t) = \chi_{\beta\alpha}^*(-t)$. In the Markovian
limit, it becomes
\begin{equation}
  \label{eq:cfr}
  \chi_{\alpha\beta}(t) \approx \frac{1}{2}\gamma_{\alpha\beta} \delta(t). 
\end{equation}
Then, we get
\begin{eqnarray}
  \label{eq:colm}
  {\cal C}^{(2)}(t) {\hat \rho}(t) &\approx& \frac{1}{2} \sum_{\alpha\beta}
  \gamma_{\alpha\beta} \{[{\hat v}_\alpha {\hat \rho}(t), {\hat
  v}_\beta] + [{\hat v}_\alpha, {\hat \rho}(t){\hat v}_\beta]\}
\end{eqnarray}
where $\gamma_{\alpha\beta}$ contains the information about the
physical decoherence parameters. It is now obvious that
Eq.~(\ref{eq:colm}) is equivalent to the Lindblad term ${\cal L}_D$
described in Ref.~\cite{Lidar98}, which takes into account the
nonunitary, decohering dynamics. 

We now proceed to prove that the OSR or the Kraus representation can
be derived from the formal solution given in Eqs.~(\ref{eq:rho1}) and
(\ref{eq:weq}). The evolution superoperator ${\cal E}(t)$ becomes
\begin{eqnarray}
\label{eq:Mequation}
{\cal {\cal E}}^{(2)}(t)&=&\left\{
1-i\int^t_0 ds ~{\cal U}_s(t,s){\rm Tr}_b\Big[
{\cal L}_{int}{\cal Z}^{(1)}(s){\hat \rho}_b\Big]
{\cal U}_s^{-1}(t,s)\right\}\nonumber \\
& & \times {\cal U}_s(t,0)
\end{eqnarray}
with
\begin{eqnarray}
{\cal Z}^{(1)}(s)=-i\int^s_0 d\tau ~{\cal U}_0(s,\tau){\cal
  L}_{int}{\cal U}_0^{-1}(s,\tau), 
\label{sigma1}
\end{eqnarray}
within the Born approximation.
Substituting Eqs.~(\ref{eq:cli}) and (\ref{eq:chif}) into
Eq.~(\ref{eq:Mequation}), Eq.~(\ref{eq:rho1}) becomes
\begin{eqnarray}
  \label{eq:sotevol}
  {\hat \rho}(t) &=& {\cal U}_s(t,0) {\hat \rho}(0) - {\cal U}_s(t,0)
  \sum_{\alpha\beta}\int_0^t ds \int_0^s d\tau \nonumber \\
  & & \times \Bigg\{ \chi_{\alpha\beta}(\tau-s) \big[ {\hat \rho}(0)
  {\hat v}_\beta(\tau) {\hat v}_\alpha(s) - {\hat v}_\alpha(s){\hat
  \rho}(0){\hat v}_\beta(\tau)\big] \nonumber \\
  & & +
  \chi_{\alpha\beta}^*(\tau-s) \big[
  {\hat v}_\alpha(s){\hat v}_\beta(\tau){\hat \rho}(0) - {\hat
  v}_\beta(\tau) {\hat \rho}(0) {\hat v}_\alpha(s) 
  \big] \Bigg\}
\end{eqnarray}

The superoperator ${\cal E}^{(2)}(t)$ satisfies the following
conditions: (i) trace-preserving, (ii) Hermiticity-preserving, and
(iii) complete positivity. As a result, there exists a corresponding
OSR \cite{Kraus71}.  We will find the OSR for ${\cal E}^{(2)}(t)$ in
Eq.~(\ref{eq:Mequation}) although any order of perturbation is
applicable based on our formulation.  Let $\{{\hat K}_\alpha\}$ be the
set of Kraus operators for ${\hat \rho}(t)$ described in
Eq.~(\ref{eq:sotevol}), then
\begin{eqnarray}
  \label{eq:osr}
  {\hat \rho}(t) = {\cal E}^{(2)}(t){\hat \rho}(0) = \sum_\alpha
  {\hat K}_\alpha(t) {\hat \rho}(0) 
  {\hat K}^\dagger_\alpha(t) 
\end{eqnarray}
with the completeness relation, independent of the evolving time
$t$,
\begin{eqnarray}
  \label{eq:osrcr}
  \sum_\alpha {\hat K}^\dagger_\alpha(t) {\hat K}_\alpha(t) = \openone.
\end{eqnarray}
In order to derive explicit expressions for the superoperator ${\cal
  E}^{(2)}$, we employ the interaction picture for the time evolution of the
system state as
\begin{eqnarray}
  \label{eq:qsip}
  {\tilde \rho}(t) &=& {\tilde {\cal E}}^{(2)}(t){\hat \rho}(0)
  \nonumber \\ 
  &=& {\cal U}_s^{-1}(t,0){\cal E}^{(2)}(t){\hat \rho}(0) \nonumber \\
  &=& \sum_\alpha {\tilde K}_\alpha(t) {\hat \rho}(0) {\tilde
  K}^\dagger_\alpha(t). 
\end{eqnarray}
To derive the set of Kraus operators $\{{\tilde K}_\alpha\}$, we adopt
a matrix representation for them. Then,
\begin{eqnarray}
  \label{eq:mrso}
  {\tilde {\cal E}}^{(2)}{\hat e}_{nm} = \sum_{ab} {\cal
  E}^{ab}_{nm} {\hat e}_{ab}, 
\end{eqnarray}
with
\begin{eqnarray}
  \label{eq:ipso}
  {\cal E}^{ab}_{nm} = ({\hat e}_{ab}, {\tilde {\cal E}}^{(2)}{\hat e}_{nm})
\end{eqnarray}
where $\{{\hat e}_{ab} | {\hat e}_{ab} = |a \rangle\langle b|\}$ is an
orthonormal basis set which spans the Hilbert-Schmidt space of reduced
density operators. The Kraus operator is expanded in this basis as
\begin{eqnarray}
  \label{eq:mrko}
  {\tilde K}_\alpha = \sum_{ab} \kappa^{ab}_\alpha {\hat e}_{ab},
\end{eqnarray}
then,
\begin{eqnarray}
  \label{eq:mrwithko}
  \sum_\alpha {\tilde K}_\alpha {\hat e}_{nm} {\tilde K}^\dagger_\alpha
  = \sum_{\alpha a b} \kappa^{an}_\alpha {\kappa^{bm}_\alpha}^*
  {\hat e}_{ab}. 
\end{eqnarray}
Comparing Eqs.~(\ref{eq:mrso}) and (\ref{eq:mrwithko}), we obtain
\begin{eqnarray}
  \label{eq:mrcom}
  {\cal E}^{ab}_{nm} = \sum_\alpha \kappa^{an}_\alpha
  {\kappa^{bm}_\alpha}^*. 
\end{eqnarray}
The conversion to ${\cal E}^{(2)}$ is straightforward since ${\cal
  U}_s(t)$ is unitary. From Eqs.~(\ref{eq:sotevol}) and
(\ref{eq:ipso}), we get
\begin{eqnarray}
  \label{eq:mrsonu}
  {\cal E}^{ab}_{nm} &=& \delta_{an}\delta_{bm} - B_{an} \delta_{bm} -
  \delta_{an} B_{bm}^* + A_{an,bm}  
\end{eqnarray}
where $\delta_{an}$ is a Kronecker delta,
\begin{eqnarray}
  \label{eq:bf}
  B_{an} &=& \sum_{\alpha\beta} \int_0^t ds \int_0^s d\tau
  ~\chi_{\alpha\beta}^*(\tau-s) \langle a|
  {\hat v}_\alpha(s){\hat v}_\beta(\tau)|n\rangle, \\
  \label{eq:af}
  A_{an,bm} &=& \sum_{\alpha\beta} \int_0^t ds \int_0^s d\tau
  ~\chi_{\alpha\beta}(\tau-s) \nonumber \\
  & & \times \langle a|{\hat v}_\alpha(s)|n\rangle\langle b| {\hat v}_\beta(\tau)|m\rangle^*.
\end{eqnarray}

The set of Kraus operators are not unique and can be generated from a
canonical set by an extended unitary matrix
\cite{Kraus71,Hughston93,Schumacher96,Nielsen00}. We will obtain the
canonical set of Kraus operators. The superoperator ${\cal
  E}_{nm}^{ab}$ can be regarded as a positive and Hermitian matrix
with $(a,n)$ being the row index and $(b,m)$ the column index
\cite{Schumacher96}. Then, there exists some unitary matrix
$U_{an,\alpha}$ which diagonalizes ${\cal E}_{nm}^{ab}$ as
\begin{eqnarray}
  \label{eq:sodiag}
  {\cal E}_{nm}^{ab} = \sum_\alpha U_{an,\alpha} d_{\alpha} U^*_{bm,\alpha}.
\end{eqnarray}
Since all eigenvalues $d_{\alpha}$ are positive, $d_{\alpha} =
\sqrt{d_{\alpha}}\sqrt{d_{\alpha}}$, and Eq.~(\ref{eq:sodiag}) is in
the form of Eq.~(\ref{eq:mrcom}). One may choose $\kappa^{an}_\alpha$
as
\begin{eqnarray}
  \label{eq:chko}
  \kappa^{an}_\alpha = \sqrt{d_\alpha} U_{an,\alpha}.
\end{eqnarray}
All equivalent sets of Kraus operators for the given superoperator can
be generated by ``unitary remixing'' of the canonical set with the
eigenvalue vector ${\bf d}'$ extended by some arbitrary number of
zeros as ${\bf d}'=({\bf d},0,...,0)$ \cite{Hughston93}.

In addition to the derivation from the canonical set, when the
superoperator is already in the form of Eq.~(\ref{eq:mrcom}), the
Kraus operators can be obtained more explicitly. As an example, let us
consider a simple dephasing channel for a single qubit system where
Hamiltonian is given by \cite{Palma97}
\begin{eqnarray}
  \label{eq:sqsham}
  {\hat H}_t 
  = \frac{1}{2} \epsilon_0 {\hat \sigma}_z + \sum_k \omega_k {\hat
  a}^\dagger_k {\hat a}_k + {\hat \sigma}_z {\hat b} 
\end{eqnarray}
where $\hat{b} = \sum_k g_k {\hat a}^\dagger_k + g^*_k {\hat a}_k$.
From Eqs.~(\ref{eq:chif}), (\ref{eq:bf}), and (\ref{eq:af}), we obtain
\begin{eqnarray}
  \label{eq:sqschif}
  \chi(t) &=& {\rm Tr}_b ({\hat b}(t) {\hat b} {\hat \rho}_b) \\
  B_{an} &=& \delta_{an} f(t), \\
  A_{an,bm} &=& 2 {\rm Re}f(t) \lambda_a \lambda_b \delta_{an} \delta_{bm}
\end{eqnarray}
where 
$\lambda_a$ is an eigenvalue of ${\hat \sigma}_z$, and
$f(t) = \int_0^t ds \int_0^s d\tau ~\chi^*(\tau-s)$. Note that the
eigenvectors $|n\rangle$ of ${\hat \sigma}_z$ was used for the basis
to obtain $B$ and $A$. If we set $\kappa^{an}_0 = \delta_{an} [1-
f(t)]$ and $\kappa^{an}_1 = \sqrt{2 {\rm Re} f(t)-|f(t)|^2} \lambda_a
\delta_{an}$, then the resulting Kraus operators are
\begin{eqnarray}
  \label{eq:sqsosr}
  {\tilde K}_0 &=& [1-f(t)] \left(
    \begin{array}{cc}
      1 & 0 \\
      0 & 1 \\
    \end{array}\right) \nonumber \\
  {\tilde K}_1 &=& \sqrt{2{\rm Re}f(t)-|f(t)|^2} \left(
    \begin{array}{cc}
      1 & 0 \\
      0 & -1 \\
    \end{array}\right).
\end{eqnarray}
If we define $p(t) = 2 {\rm Re}f(t)-|f(t)|^2$, then the OSR of ${\hat
  \rho}(t)$ becomes
\begin{eqnarray}
  \label{eq:sqsdep}
  {\hat \rho}(t) = \left(1-p(t)\right) \openone {\hat \rho}(0) \openone + p(t)
  {\hat \sigma}_z {\hat \rho}(0) {\hat \sigma}_z. 
\end{eqnarray}
It is obvious that the Kraus operators (\ref{eq:sqsosr}) satisfy the
completeness relation (\ref{eq:osrcr}).

In summary, we have shown that the OSR for the non-Markovian case and
the  Lindblad master equation for the Markov case can be
derived from the formal solution to the QLE for the qubit system in
the presence of decoherence processes self-consistently. Our
formulation is the first principle theory starting from the exact
solution to the QLE in time-convolutionless form and the matrix
representation of the evolution superoperator. 
The advantage of our first principle theory is that it is general
enough to model a realistic quantum computer in the presence of
decoherence provided that details of the Hamiltonians for the system,
reservoir, and the mutual interaction are known.

\acknowledgements

This work was supported by the Korean Ministry of Science and
Technology through the Creative Research Initiatives Program under
Contract No. 00-C-CT-01-C-35.

\end{document}